\documentclass[prb,aps,showpacs,twocolumn]{revtex4}

\usepackage{amsmath}
\usepackage{bm}
\usepackage{graphicx}

\begin{document}

\title{Dissipative structures of quantized vortices in a coherently pumped
polariton superfluid}

\author{Tomohiko Aioi}
\affiliation{Department of Engineering Science, University of
Electro-Communications, Tokyo 182-8585, Japan}

\author{Tsuyoshi Kadokura}
\affiliation{Department of Engineering Science, University of
Electro-Communications, Tokyo 182-8585, Japan}

\author{Hiroki Saito}
\affiliation{Department of Engineering Science, University of
Electro-Communications, Tokyo 182-8585, Japan}

\date{\today}

\begin{abstract}
Steady states reached in a coherently pumped exciton--polariton superfluid
are investigated.
As the pump parameter is changed, the translational symmetry of the
uniform system is spontaneously broken, and various steady patterns of
quantized vortices are formed.
This is peculiar to nonequilibrium dissipative systems with continuous
pumping.
\end{abstract}


\maketitle

\section{Introduction}

Quantum degenerate exciton--polaritons~\cite{Deng,Kasprzak,Chris} in a
semiconductor microcavity have recently attracted growing interest.
Such a system consists of quantum wells sandwiched between two distributed
Bragg reflectors, where excitons in the quantum wells are coupled with
photons confined in the reflectors.
The coupling of the excitons with photons makes the exciton--polariton
mass much smaller than the electron mass, which enables the system to
maintain long-range coherence.
In recent years, experimental studies on this system have shown rapid
progress~\cite{Review}: for example, Bogoliubov
excitations~\cite{Utsunomiya} and quantized
vortices~\cite{Lagoudakis,Lagoudakis09,Roumpos} have been observed, and
various quantum fluid dynamics have been
demonstrated.~\cite{Amo09N,Amo09,Amo11,Nardin,Sanvitto,Grosso}
In the present paper, we focus on a system in which polaritons are
continuously pumped by near-resonant laser beams.~\cite{Amo09,Amo11}
The loss of polaritons with a short lifetime is balanced by the continuous
pumping, thereby realizing a nonequilibrium open system.

Homogeneous steady states far from equilibrium can spontaneously break the
translational symmetry, and they can exhibit patterns or structures,
called dissipative structures.~\cite{Nicolis}
This kind of structure formation occurs in a variety of systems, such as
Rayleigh--B\'enard convection cells~\cite{Benard} in a fluid heated from
below, and Belousov--Zhabotinsky patterns~\cite{Zaikin} and Turing
patterns~\cite{Ouyang} in chemical reaction--diffusion systems.
However, few studies have been performed on dissipative structures in
quantum mechanical systems.~\cite{Lugiato,Chernyuk}
The polariton superfluid in a microcavity is a suitable system for
realizing quantum dissipative structures, since polaritons can be injected
into the system continuously and coherently with relative
ease.~\cite{Review}
The continuous injection and dissipation of polaritons is expected to
drive the formation of quantum dissipative structures.

\begin{figure}[tbp]
\includegraphics[width=8cm]{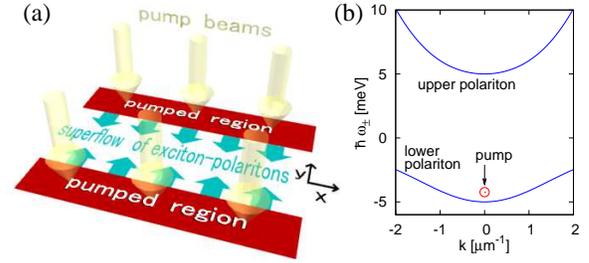}
\caption{
(color online) (a) Schematic illustration of the system.
The quasi-two-dimensional system lies in the $x$-$y$ plane and is pumped
by laser beams applied from the $z$ direction.
This coherently excites the polaritons with zero momentum.
The pumped regions are restricted to $|y| > d / 2$ and the pumped
polariton superfluid flows into the interspace $|y| < d / 2$.
(b) Dispersion relation of the lower and upper polaritons.
The pump energy and momentum correspond to the circle.
}
\label{f:intro}
\end{figure}
In the present paper, we show that various steady patterns of quantized
vortices are formed in an exciton--polariton superfluid in a semiconductor
microcavity pumped by near-resonant continuous-wave laser beams.
We consider a situation in which two spatial regions are pumped and a
polariton superfluid flows into the interspace between the two pumped
regions, as illustrated in Fig.~\ref{f:intro} (a).
When the interval $d$ between the two regions is small, the steady state
has continuous translational symmetry (with respect to $x$ in
Fig.~\ref{f:intro} (a)).
When $d$ exceeds a critical value, an instability sets in and the
continuous translational symmetry is spontaneously broken.
The system develops into another steady state, in which quantized vortices
are aligned periodically.
These patterns of quantized vortices are spontaneously organized in
nonequilibrium steady states and are peculiar to the open system with loss
and pump.
We also show that the pattern depends on $d$ and the phase difference
between the two pumped regions.
In finite systems, vortices escape from the edges of the pumped regions,
which can be plugged by an appropriate potential.

This paper is organized as follows.
Section~\ref{s:form} provides the formulation of the problem.
Section~\ref{s:num} numerically demonstrates the formation of various
patterns.
Section~\ref{s:conc} presents the conclusions to this study.

\section{Formulation of the problem}
\label{s:form}

We used the mean-field theory to study the dynamics of an
exciton--polariton superfluid.
The energy of a photon in a quasi-two-dimensional microcavity can be
approximated as $\hbar \omega_{\rm C}^0 + p^2 / (2 M_{\rm C})$, where
$\bm{p}$ is the in-plane momentum and $M_{\rm C}$ is the effective mass of
a photon.
The effective mass of an exciton is much larger than $M_{\rm C}$, and we
approximated the energy of an exciton as a constant,
$\hbar \omega_{\rm X}^0$.
The excitons in the quantum wells and the photons in the microcavity
couple with the Rabi frequency $\Omega_{\rm R}$.
The macroscopic wave functions for the photons $\psi_{\rm C}$ and the
excitons $\psi_{\rm X}$ are assumed to obey the coupled nonlinear
Schr\"odinger equations given by~\cite{Carusotto}
\begin{subequations} \label{GP}
\begin{eqnarray}
\label{GP1}
i \hbar \frac{\partial \psi_{\rm C}}{\partial t} & = &
\left( \hbar \omega_{\rm C}^0 - \frac{\hbar^2}{2 M_{\rm C}} \nabla^2
\right) \psi_{\rm C} + \hbar \Omega_{\rm R} \psi_{\rm X} + V(\bm{r})
\psi_{\rm C} \nonumber \\
& & - i \frac{\hbar \gamma_{\rm C}}{2} \psi_{\rm C}
+ e^{i (\bm{k}_{\rm p} \cdot \bm{r} - \omega_{\rm p} t)} F(\bm{r}), \\
i \hbar \frac{\partial \psi_{\rm X}}{\partial t} & = &
\hbar \omega_{\rm X}^0 \psi_{\rm X} + \hbar \Omega_{\rm R} \psi_{\rm C}
+ g |\psi_{\rm X}|^2 \psi_{\rm X}
- i \frac{\hbar \gamma_{\rm X}}{2} \psi_{\rm X},
\end{eqnarray}
\end{subequations}
where $V(\bm{r})$ is the external potential for the photons, $g$ is an
exciton--exciton interaction coefficient, and $\gamma_{\rm C (X)}$ is the
decay constant for the photons (excitons).
The last term on the right-hand side of Eq.~(\ref{GP1}) describes the
coherent pumping of photons by external laser beams, where
$\bm{k}_{\rm p}$, $\omega_{\rm p}$, and $F(\bm{r})$ are the in-plane wave
number, frequency, and profile of the pumping, respectively.

Diagonalizing the first and second terms on the right-hand side of
Eq.~(\ref{GP}) for the mode with wave number $\bm{k}$, we obtain the
energies of the upper and lower polaritons as
\begin{equation} \label{disp}
\omega_\pm(k) = \frac{1}{2} \left[ \omega_k^0 + \omega_{\rm C}^0 +
\omega_{\rm X}^0 \pm \sqrt{(\omega_k^0 + \omega_{\rm C}^0
- \omega_{\rm X}^0)^2 + 4 \Omega_{\rm R}^2} \right], 
\end{equation}
where $\omega_k^0 = \hbar k^2 / (2 M_{\rm C})$.
The dispersion relation in Eq.~(\ref{disp}) is depicted in
Fig.~\ref{f:intro}(b).
The difference between the pump and lower polariton frequencies is defined
as
\begin{equation}
\delta = \omega_{\rm p} - \omega_-(0).
\end{equation}
For simplicity, in the following calculations, we restrict ourselves to
the case of $\omega_{\rm C}^0 = \omega_{\rm X}^0 = 0$, $\gamma_{\rm C} =
\gamma_{\rm X} \equiv \gamma$, and $\bm{k}_{\rm p} = 0$.
The parameters are fixed as follows: $M_{\rm C} = 2 \times 10^{-5}
m_{\rm e}$ with $m_{\rm e}$ being the electron mass, $\hbar \Omega_{\rm R}
= 5$ meV, $g = 0.01$ ${\rm meV}\mu{\rm m}^2$, $\gamma^{-1} = 10$ ps, and
$\delta = 0.76$ meV.

\section{Numerical results}
\label{s:num}

\subsection{Steady states with continuous translational symmetry}

We first consider an infinite system without an external potential ($V =
0$).
We use a pump profile as illustrated in Fig.~\ref{f:intro}(a), given by
\begin{equation} \label{pump}
F(\bm{r}) = \left\{ \begin{array}{ll}
F_0 & (y \leq -d / 2), \\
0 & (|y| < d / 2), \\
F_0 e^{i \phi} & (y \geq d / 2),
 \end{array} \right.
\end{equation}
where $F_0 = 5.7$ meV, $\phi$ is the phase difference between the two pump
beams, and $d$ is the width of the interspace between the two pumped
regions.
For this value of $F_0$, the pumped regions are excited into the stable
uppermost branch of the bistability curve.~\cite{Carusotto}

\begin{figure}[tbp]
\includegraphics[width=8cm]{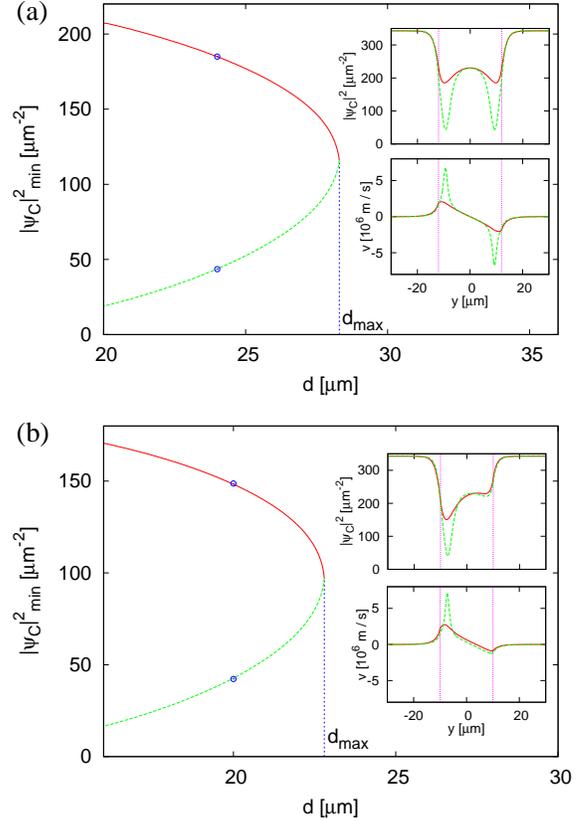}
\caption{
(color online) Steady states having continuous translational symmetry with
respect to $x$, which are obtained by solving Eq.~(\ref{GP}) with pumping
as in Eq.~(\ref{pump}).
The phase difference in the pump is $\phi = 0$ in (a) and $\phi = \pi$ in
(b).
The main panels show the minimum value of $|\psi_{\rm C}(y)|^2$ of the
steady state (corresponding to the density dips as shown in the inset).
For $d < d_{\rm max}$, there exist stable (red solid) and unstable (green
dashed) steady states.
The density and velocity distributions of the photonic states
corresponding to the circles are shown in the insets, where the red solid
and green dashed curves indicate the stable and unstable steady states.
The vertical lines in the insets indicate $\pm d / 2$.
}
\label{f:saddle}
\end{figure}
In the limit of $d \rightarrow 0$ with $\phi = 0$, the whole system is
uniformly pumped into the stable state $\psi_{\rm C, X}(\bm{r}, t) \propto
e^{-i \omega_{\rm p} t}$.
When $d$ is increased, the density at $|y| \lesssim d / 2$ decreases and
the continuous translational symmetry with respect to $x$ is maintained if
$d$ is not very large.
We sought such steady states that is uniform in $x$ by solving
Eq.~(\ref{GP}) using the Newton-Raphson method.
We also checked the stability of the steady states by using Bogoliubov
analysis.
If there is a Bogoliubov frequency that has a positive imaginary part, the
state is dynamically unstable.
Figure~\ref{f:saddle} shows the $d$-dependence of the steady states.
For $\phi = 0$, the steady states have even-parity density profiles
with two minima, where the velocity is largest, as shown in the inset in
Fig.~\ref{f:saddle}(a)
(the exciton wave function $\psi_{\rm X}$ is always qualitatively the same
as the photon wave function $\psi_{\rm C}$, and thus in the following
discussion, we will show only $\psi_{\rm C}$).
There are two steady states for each $d$;
one is stable (red solid curve) and the other is dynamically unstable
(green dashed curve).
The dynamically unstable modes grow exponentially and break the
continuous translational symmetry with respect to $x$.
The density dips in the unstable state are deeper than those of the stable
state.
These minimum values of the density are shown in the main panels of
Fig.~\ref{f:saddle}.
As $d$ is increased, the stable and unstable branches meet at
$d = d_{\rm max}$ and the steady states disappear for $d > d_{\rm max}$,
which is a manifestation of the saddle-node bifurcation.
Figure~\ref{f:saddle}(b) shows the case of $\phi = \pi$.
The $d$-dependence is similar to that in Fig.~\ref{f:saddle}(a) except
that the density and velocity profiles are asymmetric, as shown in the
inset of Fig.~\ref{f:saddle}(b).
There are two degenerate states, since Eq.~(\ref{GP}) is symmetric with
respect to $y \rightarrow -y$ and $\psi_{\rm C, X} \rightarrow
-\psi_{\rm C, X}$ for $\phi = \pi$ (only one state is plotted in the
inset).

From Fig.~\ref{f:saddle}, we expect the following scenario:
When the initial value of $d$ is small and then $d$ is increased
adiabatically, the system follows the stable steady states (red solid
curve in the main panels of Fig.~\ref{f:saddle}).
When $d$ exceeds $d_{\rm max}$, the steady states disappear, which causes
the breaking of the continuous translational symmetry with respect to
$x$.
The system then goes to another nonequilibrium steady state with broken
symmetry, if it exists.
In the next subsection, we will show that these states have a variety of
patterns with quantized vortices.
This scenario of pattern formation is similar to that in the dissipative
structures~\cite{Nicolis}.

The disappearance of the steady states for a large $d$ can be understood
as follows.
The steady states have the factor of $e^{-i \omega_{\rm p} t}$ and
therefore have the ``energy'' of $\hbar \omega_{\rm p}$.
Since the kinetic energy of the superflow must be smaller than this energy
$\hbar \omega_{\rm p}$, the flow velocity has an upper bound
$v_{\rm max}$.
The loss of polaritons per unit time at $y < |d| / 2$ is $\sim \gamma \rho
d$, where $\rho$ is the polariton density.
The loss must be replenished by injection $\sim \rho v$, i.e., $\gamma
\rho d \sim \rho v$.
Thus, $d$ has an upper bound $d_{\rm max} \sim v_{\rm max} / \gamma$.
We have numerically confirmed that $d_{\rm max}$ is roughly proportional
to $\gamma^{-1}$.

\subsection{Symmetry breaking and pattern formations}

\begin{figure}[tb]
\includegraphics[width=7.5cm]{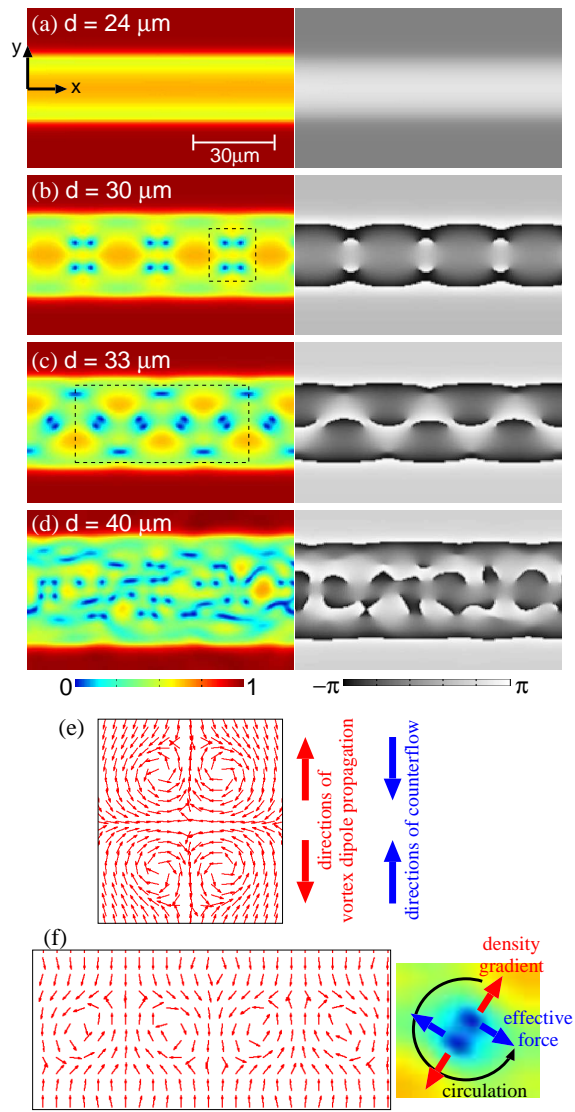}
\caption{
(color online) (a)--(d) Density ($|\psi_{\rm C}|^2$, left panels) and
phase (${\rm arg} \psi_{\rm C}$, right panels) profiles of the steady
states for (a) $d = 24$ $\mu{\rm m}$, (b) $d = 30$ $\mu{\rm m}$, (c)
$d = 33$ $\mu{\rm m}$, and a snapshot of an unsteady state for (d)
$d = 40$ $\mu{\rm m}$. 
The unit of the density is 350 $\mu{\rm m}^{-2}$.
The flow fields in the dashed squares in (b) and (c) are shown in (e) and
(f), respectively.
The two pumped regions have the same phase, $\phi = 0$, which corresponds
to Fig.~\ref{f:saddle}(a).
See Supplemental Material~\cite{movies} for the dynamics, in which $d$ is
changed from $24$ $\mu{\rm m}$ to $40$ $\mu{\rm m}$.
}
\label{f:main}
\end{figure}
Figures~\ref{f:main}(a)--\ref{f:main}(c) show the stable steady states for
$\phi = 0$.
The steady state shown in Fig.~\ref{f:main}(a) is uniform in $x$, which
corresponds to the stable branch for $d < d_{\rm max}$ in
Fig.~\ref{f:saddle}(a).
As $d$ increases and exceeds $d_{\rm max}$, the stable branch in
Fig.~\ref{f:saddle}(a) disappears.
As a consequence, the continuous translational symmetry with respect to
$x$ is spontaneously broken and quantized vortex pairs are created;
these align periodically, as shown in Fig.~\ref{f:main}(b).
The vortex state shown in Fig.~\ref{f:main}(b) is a stable steady state,
in which vortex dipoles (clockwise and counterclockwise vortex pairs) are
the building blocks of the pattern: two vortex dipoles form a pair that
has parity symmetry with respect to $y$.
Although a vortex dipole propagates at a constant velocity in a uniform
system, the vortex dipoles in Fig.~\ref{f:main}(b) remain at rest.
This is because the vortex dipoles are prevented from propagating by the
counterflow from the pumped regions, as shown in Fig.~\ref{f:main}(e).
The polaritons pumped in the region of $y > d / 2$ ($y < -d / 2$) flow
in the direction of $-y$ ($+y$) and sustain the vortex dipoles located in
$y > 0$ ($y < 0$).
These vortex dipoles would propagate in the $+y$ ($-y$) direction if they
were in a uniform system without superflow.

As $d$ is increased further, the steady vortex structure in
Fig.~\ref{f:main}(b) makes a transition to another vortex structure, as
shown in Fig.~\ref{f:main}(c).
The vortex pairs aligned along $y \simeq 0$ are not vortex dipoles but
twin vortices in which two vortices have the same circulation.
Such twin vortices with clockwise and counterclockwise circulations
alternate along $y \simeq 0$, resulting in a structure similar to the
vortex street reported in Ref.~\onlinecite{Sasaki}.
Although two vortices with the same circulation rotate around one another
in a uniform system, the twin vortices in Fig.~\ref{f:main}(c) are at
rest.
This can be understood from the Magnus force on the vortices, which is due
to the density distribution~\cite{Aioi} in Fig.~\ref{f:main}(c):
a vortex with circulation $\bm{\kappa}$ experiences an effective force
proportional to $-\bm{\kappa} \times \bm{\nabla} \rho$, where
$\rho$ is the polariton density.
For example, in a uniform system, the twin vortices magnified in the right
panel of Fig.~\ref{f:main}(f) would rotate around one another in the
counterclockwise direction.
The density gradient in the directions of the red (light gray) arrows
exert forces in the directions of the blue (dark gray) arrows, which holds
the rotation of the twin vortices.

The continuous translational symmetry with respect to $x$, as shown in
Fig.~\ref{f:main}(a), thus changes to discrete translational symmetry due
to the vortex pattern formation, as shown in Figs.~\ref{f:main}(b) and
\ref{f:main}(c).
For a larger $d$, there is no steady state, and the vortex pattern
dynamically changes in a random manner, which has no symmetry
(a snapshot of the dynamics is shown in Fig.~\ref{f:main}(d)).

\begin{figure}[tb]
\includegraphics[width=8cm]{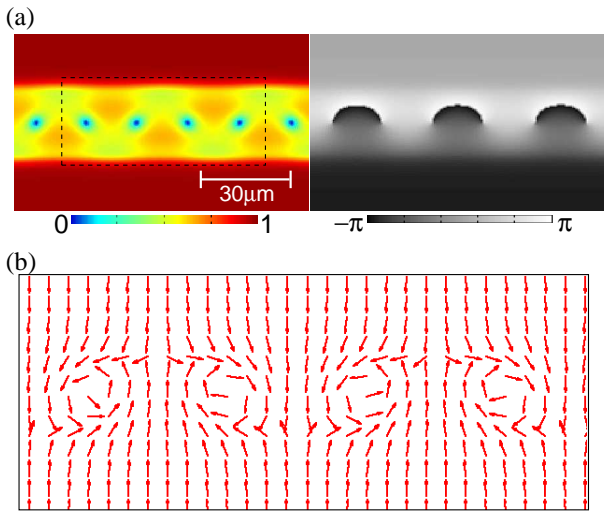}
\caption{
(color online) (a) Density ($|\psi_{\rm C}|^2$) and phase (${\rm arg}
\psi_{\rm C}$) profiles of the steady state for $d = 24$ $\mu{\rm m}$,
where the phase difference between the two pumped regions is $\phi = \pi$.
The unit of density is 350 $\mu{\rm m}^{-2}$.
(b) The flow field in the dashed square in (a).
See Supplemental Material~\cite{movies} for the dynamics in which $\phi$
is changed from $0$ to $\pi$.
}
\label{f:piphase}
\end{figure}
Figure~\ref{f:piphase} shows the steady state for $\phi = \pi$, which
corresponds to the case of Fig.~\ref{f:saddle}(b).
For the width $d = 24$ $\mu{\rm m}$, there is no steady state that is
uniform in $x$, and a vortex pattern emerges as shown in
Fig.~\ref{f:piphase}.
The clockwise and counterclockwise vortices are aligned alternately along
$y \simeq 0$, whose flow pattern (Fig.~\ref{f:piphase}(b)) is similar to
that in Fig.~\ref{f:main}(f).
For a classical incompressible inviscid fluid, such an arrangement of
point vortices is unstable.
(The only stable arrangement of this kind of vortex rows is the one found
by von K\'arm\'an.~\cite{Lamb})
The vortex pattern in Fig.~\ref{f:piphase} is stabilized by the superflow
from the pumped regions $|y| > d / 2$ toward $y = 0$.
The density distribution around the vortices may also be responsible for
the stabilization.

The vortex patterns in Figs.~\ref{f:main}(b), \ref{f:main}(c), and
\ref{f:piphase} are periodic in $x$, and their periods are $\simeq 28$
$\mu{\rm m}$, $31$ $\mu {\rm m}$, and $34$ $\mu{\rm m}$, respectively.
The periods of the vortex patterns are determined by the geometry of the
system, and they are of the order of the interval $d$ between the pumped
regions.

\subsection{Finite systems}

\begin{figure}[tb]
\includegraphics[width=7cm]{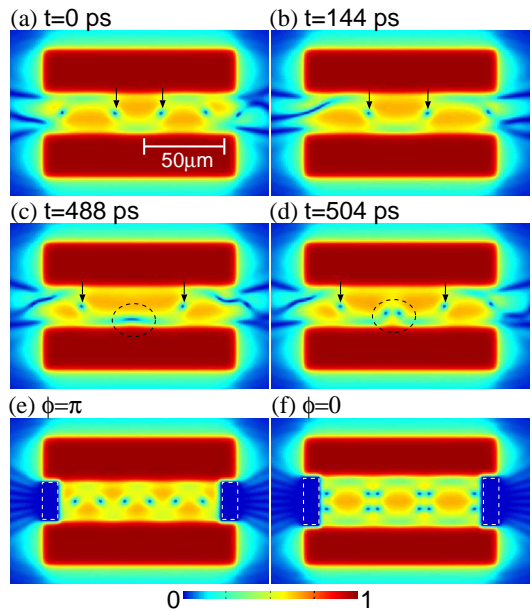}
\caption{
(color online) Density profiles ($|\psi_{\rm C}|^2$) for the finite pumped
regions given by Eq.~(\ref{finite}).
(a)--(d) Snapshots of the dynamics for $\phi = \pi$ and
$d = 24$ $\mu{\rm m}$.
The origin for the time is chosen appropriately.
The arrows trace the vortex motion.
The dashed circles indicate a created vortex pair.
(e), (f) Steady states with the plug potentials given by Eq.~(\ref{pot})
with $V_0 = 380$ meV for (e) $\phi = \pi$ and $d = 30$ $\mu{\rm m}$, and
(f) $\phi = 0$ and $d = 24$ $\mu{\rm m}$.
The plug potentials are marked by the dashed squares.
The unit of the density is 350 $\mu{\rm m}^{-2}$.
See Supplemental Material~\cite{movies} for the dynamics in (a)--(d).
}
\label{f:finite}
\end{figure}
Thus far, we have considered only infinite systems, in which the steady
states have continuous or discrete translational symmetry with respect to
$x$.
In this section, we examine pumped regions of finite size, as follows:
\begin{equation} \label{finite}
F(\bm{r}) = \left\{ \begin{array}{ll}
F_0 & (|x| \leq 60 \mbox{$\mu {\rm m}$},
-40 \mbox{$\mu {\rm m}$} \leq y \leq -d / 2), \\
F_0 e^{i \phi} & (|x| \leq 60 \mbox{$\mu {\rm m}$},
d / 2 \leq y \leq 40 \mbox{$\mu {\rm m}$}), \\
0 & (\mbox{otherwise}),
\end{array} \right.
\end{equation}
where $F_0 = 5.7$ meV is the same as that in Figs.~\ref{f:main} and
\ref{f:piphase}.
Figures~\ref{f:finite}(a)--\ref{f:finite}(d) show the dynamics for the
parameters $\phi = \pi$ and $d = 24$ $\mu{\rm m}$, where the origin of
time is chosen appropriately.
For these values of $d$ and $\phi$, the infinite pumped system has the
steady state with a vortex pattern, as shown in Fig.~\ref{f:piphase}.
In the finite system, however, the polaritons in the space between the
pumped regions flow out from the edges at $|x| = 60$ $\mu{\rm m}$.
As a consequence, vortices also flow out, as indicated by the arrows in
Figs.~\ref{f:finite}(a)--\ref{f:finite}(d).
As the vortices flow out, they separate from each other, and new
vortex-antivortex pairs are created between them, as shown by the dashed
circles in Figs.~\ref{f:finite}(c) and \ref{f:finite}(d).
Such separations and creations of vortices are repeated, and the system
does not reach a steady state.

To stabilize the system and obtain steady states, we add a plug potential
as
\begin{equation} \label{pot}
V(\bm{r}) = \left\{ \begin{array}{ll}
V_0 & (50 \mbox{$\mu {\rm m}$} \leq |x| \leq 60 \mbox{$\mu {\rm m}$},
|y| \leq -d / 2), \\
0 & (\mbox{otherwise}),
\end{array} \right.
\end{equation}
which prevents the polaritons from flowing out.
Figures~\ref{f:finite}(e) and \ref{f:finite}(f) show the steady states for
$\phi = \pi$ and $\phi = 0$, where the regions of $V = V_0$ are enclosed
by the dashed squares.
The clockwise and counterclockwise vortices alternately align in
Fig.~\ref{f:finite}(e) and the vortex dipoles make pairs in
Fig.~\ref{f:finite}(f), which are similar to Figs.~\ref{f:piphase} and
\ref{f:main}(b), respectively.
Thus, the steady states can also be obtained for finite systems by using a
plug potential that confines polaritons to a finite region.
The steady-state patterns are suitable for time-integrated imaging in
experiments.

\section{Conclusions}
\label{s:conc}

We have investigated the steady states of an exciton--polariton superfluid
coherently pumped into a semiconductor microcavity;
this is a nonequilibrium open system.
We considered the situation in which polaritons with $k = 0$ are pumped
into two regions and flow into the space between them, as illustrated
in Fig.~\ref{f:intro}.
When the distance $d$ between the two pumped regions is small, there
exists a stable steady state with continuous translational symmetry with
respect to $x$, as shown in Fig.~\ref{f:main}(a).
As $d$ is increased, this steady state disappears (Fig.~\ref{f:saddle}),
and the system goes into other steady states with periodic patterns of
quantized vortices (Figs.~\ref{f:main}(b), \ref{f:main}(c), and
\ref{f:piphase}).
Such patterns are specific to a nonequilibrium system with loss and pump,
and they can be recognized as dissipative structures.
We have shown that steady-state patterns can also be observed in a finite
system with an appropriate potential (Fig.~\ref{f:finite}).

The structure formations shown in the present paper are peculiar to
nonequilibrium steady states and should be distinguished from those in
equilibrium systems, such as crystals.
For example, the triangular lattice structure of quantized vortices in a
rotating superfluid cannot be regarded as a dissipative structure, since
it is the ground state in the rotating frame of reference.

{\it Note added.} Very recently, a preprint~\cite{Manni} has appeared,
which reports experimental and numerical observations of self-arranged
vortex--antivortex pairs in a polariton condensate.

\begin{acknowledgments}
This work was supported by Grants-in-Aid for Scientific
Research (No.\ 22340116 and No.\ 23540464) from the Ministry of Education,
Culture, Sports, Science and Technology of Japan.
\end{acknowledgments}

\end{document}